\documentstyle[12pt]{article}
\newcommand{\be}{\begin{equation}}
\newcommand{\ee}{\end{equation}}
\newcommand{\bea}{\begin{eqnarray}}
\newcommand{\eea}{\end{eqnarray}}
\newcommand{\al}{\alpha}
\newcommand{\bt}{\beta}

\newcommand{\Gm}{\Gamma}
\newcommand{\dl}{\delta}
\newcommand{\Dl}{\Delta}
\newcommand{\eps}{\epsilon}
\newcommand{\zt}{\zeta}
\newcommand{\et}{\eta}
\newcommand{\th}{\theta}

\newcommand{\kp}{\kappa}
\newcommand{\lm}{\lambda}
\newcommand{\ks}{\xi}
\newcommand{\rh}{\rho}

\newcommand{\sg}{\sigma}

\newcommand{\ta}{\tau}

\newcommand{\ph}{\phi}
\newcommand{\Ph}{\Phi}

\newcommand{\ch}{\chi}

\newcommand{\om}{\omega}
\newcommand{\Om}{\Omega}

\newcommand{\rarrow}{\rightarrow}
\newcommand{\Rarrow}{\Rightarrow}
\newcommand{\nn}{\nonumber}

\begin{document}
\title{Particle creation, renormalizability conditions and the 
mass-energy spectrum in gravity theories of quadratic Lagrangians}
\author{K Kleidis and D B Papadopoulos\\
{\small Department of Physics}\\
{\small Section of Astrophysics, Astronomy and Mechanics}\\
{\small Aristotle University of Thessaloniki}\\
{\small 54006 Thessaloniki, Greece} }

\maketitle
\begin{abstract}

Massive scalar particle production, due to the anisotropic 
evolution of a five-dimensional spacetime, is considered in 
the context of a quadratic Lagrangian theory of gravity. Those 
particles, corresponding to field modes with non-vanishing momentum 
component along the fifth dimension, are created mostly in the 
neighbourhood of a singular epoch where only their high-frequency
behaviour is of considerable importance. At the one-loop approximation 
level, general renormalizability conditions on the physical quantities 
relevant to particle production are derived and discussed. Exact 
solutions of the resulting Klein-Gordon field equation are obtained 
and the mass-energy spectrum, attributed to the scalar field due to the 
cosmological evolution, is being further investigated. Finally, analytic 
expressions regarding the number and the energy density of the created 
particles at late times, are also derived and discussed.

\end{abstract}
\section*{1. Introduction}

Gravitational Lagrangian densities, which include higher-order 
curvature terms in connection to the Einstein-Hilbert (EH) one, 
are suggested by {\em superstring theories} (Candelas et al. 1985, 
Green et al. 1987) and by the one-loop approximation of {\em quantum 
gravity} (Birrell and Davies 1982, Barrow and Ottewill 1983). 
Quadratic Lagrangians, in particular, have been used to yield 
renormalizable theories of gravity coupled to matter (Stelle 1977). 
They can also help us to improve the {\em semiclassical approximation}, 
where quantized matter fields interact with a classical gravitational 
field (Utiyama and De Witt 1962). In fact, renormalization of the 
energy-momentum tensor for a quantum field in four dimensions, 
indicates that the presence of quadratic terms in the gravitational 
action is {\em a priori} expected (Stelle 1977, 1978, Horowitz and Wald 
1978).

However, every quadratic combination of curvature terms is not 
physically accepted, since its introduction into the gravitational 
action leads to differential equations of the fourth order with 
respect to the metric (Farina-Busto 1988) and those higher-derivative 
terms are associated with {\em ghost particles} (Weinberg 1979, Zwiebach 
1985). A {\em ghosts-free}, non-linear Lagrangian theory of gravity was 
formulated by Lovelock (Lovelock 1971). He proposed that the most general 
gravitational Lagrangian is of the form
\be
{\cal L} = \sqrt {- g} \sum_{m=0}^{n/2} \lm_m {\cal L}^{(m)}
\ee
where $\lm_m$ are arbitrary constants, $n$ denotes the spacetime 
dimensions, $g$ is the determinant of the metric tensor and 
${\cal L}^{(m)}$ are functions of the Riemann curvature tensor 
${\cal R}_{\mu \nu \kp \lm}$ and its contractions ${\cal R}_{\mu \nu}$ 
and ${\cal R}$, of the form 
\be
{\cal L}^{(m)} = {1 \over 2^m} 
\dl_{\al_1...\al_{2m}}^{\bt_1...\bt_{2m}} 
{\cal R}_{\bt_1 \bt_2}^{\al_1 \al_2} ...
{\cal R}_{\bt_{2m-1} \bt_{2m}}^{\al_{2m-1} \al_{2m}}
\ee
where $\dl_{\al_1...\al_{2m}}^{\bt_1...\bt_{2m}}$ is the 
generalized Kronecker symbol. In Eq. (2),
${\cal L}^{(1)} = {1 \over 2} {\cal R}$ is the EH Lagrangian, while 
${\cal L}^{(2)}$ is a particular combination of the quadratic terms, 
known as the Gauss-Bonnett (GB) combination, which in four dimensions 
satisfies the functional relation (Kobayashi and Nomizu 1969)
\be
{\dl \over \dl g^{\mu \nu}} \int \sqrt {-g} \left ( {\cal R}^2 - 4 
{\cal R}_{\mu \nu}{\cal R}^{\mu \nu} + {\cal R}_{\mu \nu \kp \lm}
{\cal R}^{\mu \nu \kp \lm} \right ) d^4 x = 0
\ee
An important aspect of a ghosts-free, higher-order gravitational 
theory arises from the combination of Eqs. (1) and (3). Accordingly, for a 
gravitational Lagrangian with curvature terms of orders higher or equal 
than the second $(m \geq 2)$, one needs to have a spacetime of more than 
four dimensions.

The idea that the spacetime may have more than four dimensions was 
introduced by Kaluza and Klein in their efford to unify gravity and 
electromagnetism (Kaluza 1921, Klein 1926). Recently, 
higher-dimensional theories have been studied as an attractive way 
to unify all gauge interactions with gravity, in a {\em 
supergravity} scenario (Freund and Rubin 1980, Englert 1982), while 
established as unavoidable in superstring theories (Forgacs and
Horvath 1979, Green et al. 1987). In most higher-dimensional theories 
of gravity the extra dimensions are assumed to form, at the present epoch, 
a compact manifold ({\em internal space}) of very small size compared to 
that of the three-dimensional visible space ({\em external space}) 
(Applequist et al. 1987). This size is directly related to the foundamental 
constants and consequently must be stable (Accetta et al. 1986). In this 
context, it has been proposed that compactification  of the internal space 
may arise as a result of the cosmological evolution (Abbott et al. 1984, 
Kolb et al. 1984). Recent developments in this topic indicate that, contraction 
of the extra dimensions in the EH cosmology, could lead to particle 
production in the visible space (Maeda 1984a, 1984b, Garriga and Verdaguer 1989). 

In the present article we discuss creation of massive scalar particles in 
the external space, as a quantum consequence of the cosmological evolution 
in a quadratic, higher-dimensional theory of gravity. In the search for general 
renormalizability conditions regarding the quantities relevant to particle 
production, we find that in a $C^{\infty}$, globally hyperbolic, 
$n$-dimensional spacetime the total probability (P) to produce a particle pair 
over the entire history of the Universe is a positive, covariant and (what's 
most important) {\em finite} quantity, provided that the spacetime dimension 
is an {\em odd} number. The same is also true for the vacuum expectation value 
of the field's energy-momentum tensor in curved spacetime, since the 
divergences of P are the same as those that afflict $<0_{out} \vert T^{\mu \nu} 
\vert 0_{in}>$ (Hartle and Hu 1980, Birrell and Davies 1982). The most simple 
extrapolation of the four-dimensional spacetime to an odd-dimensional one is 
a curved background of five dimensions. Accordingly, as a model to perform our 
computations, we consider a five-dimensional spacetime in which the internal 
space is reduced to a static size as $t \rarrow \infty$, accompanied by an 
isotropic expansion of the visible space.

Once the appropriate background has been determined, particle production 
may be performed explicitly through the solution of the wave equation for a 
scalar field in curved spacetime under consideration. No general exact 
solutions of the equation governing the propagation of a quantum field in 
curved spacetime have been derived so far, within the context of quadratic 
theories of gravity (Mijics et al. 1985). Nevertheless, in the present paper, 
we solve explicitly the corresponding Klein-Gordon equation, to obtain exact 
massive mode solutions in terms of hypergeometric functions. Accordingly, two 
major aspects of the theory are revealed: {\bf (i)} The mass-energy spectrum, 
attributed to the scalar field due to the cosmological evolution, is discrete 
(and hence the energy is quantized), as a result of compactification of the 
extra dimension and {\bf (ii)} the energies of the produced particles are 
strongly related to the inverse radius of the internal space. Finally, using 
a high-frequency approximation, which, as we show, corresponds to creation 
of particles in the neighbourhood of a singular epoch, we arrive at analytic 
expressions regarding the {\em "observable"} number and energy density of 
the particles produced in four dimensions. 

\section*{2. The model Universe}

We look for solutions of the cosmological field equations which may 
be obtained, through Hamilton's principle, from a five-dimensional 
action with gravitational part of the form $(\hbar = 1 = c)$ (Kleidis 
and Papadopoulos 1997b) 
\be
S_{GR} = {1 \over L_5} \int \sqrt {-g} \left [ {1 \over 2 \kp} 
{\cal R} + \al \left ( {\cal R}^2 - 4 {\cal R}_{\mu \nu} 
{\cal R}^{\mu \nu} + {\cal R}_{\mu \nu \kp \lm} 
{\cal R}^{\mu \nu \kp \lm} \right ) \right ] d^5 x
\ee
In Eq. (4), $\kp = 8 \pi G$, where $G$ is the gravitational constant 
in four dimensions, $\al$ is a dimensionless constant and 
$L_5 = 2 \pi R_5 $ is a normalization constant, corresponding 
to the {\em physical size} of the internal space, once it can be 
considered static (Farina-Busto 1988). A simple cosmological solution 
with the properties required for particle production has already been 
obtained in the context of the quadratic theory under consideration (Kleidis 
and Papadopoulos 1997b). It consists of a five-dimensional spacetime with 
large scale anisotropic evolution between the two subspaces (the external 
space and the internal one). This model arises as an exact solution of the 
gravitational field equations when the minimum value of the external scale 
function is zero and the matter that fills the Universe has the following 
equation of state 
\be
p_{ext} = {1 \over 3} \rh \; \; , \; \; \; p_{int} = 0
\ee
where $p_{ext}$ and $p_{int}$ are the pressures of the external and 
the internal space, respectively and $\rh$ is the five-dimensional 
matter-energy density. The corresponding line-element is written in 
the form 
\be
ds^2 = -dt^2 + t^2 \left [ {dr^2 \over 1 + r^2} 
+ r^2 \left ( d \th^2 + \sin{\th}^2 d \ph^2 \right ) \right ] + 
\left ( 1 - {t_s^2 \over t^2} \right )^2 (dx^5)^2
\ee
where $t_s$ is a constant of dimensions $T$. The model (6) is a Ricci 
flat solution, i.e. ${\cal R} = 0$. In this solution, the hyperbolic 
$(\kp = -1)$ external space exhibits isotropic expansion, governed by the 
scale function
\be
g_{ii}(t) = R^2(t) = t^2 \:, \; \; i = 1, 2, 3
\ee
Thus, the visible Universe coincides with the asymptotic Milne phase of the 
open Friedmann-Robertson-Walker (FRW) models. As regards the internal space, 
it is assumed to be compact, i.e. closed and bounded. Therefore, at each 
time, the values of the fifth coordinate are restricted to a certain 
range
\be
0 \leq x^5 \leq 2 \pi R_5
\ee
The time evolution of the extra dimension, governed by the scale 
function
\be
g_{55}(t) = S^2(t) = \left (1 - {t_s^2 \over t^2} \right )^2
\ee
consists of an initial contraction, up to a minimum size at $t = t_s$, 
followed by a subsequent expansion (at a much lower rate) towards a 
static size, thus achieving spontaneous compactification as $t \rarrow 
\infty$ (see Fig. 1).

For $t \neq 0$ the determinant of the metric tensor under consideration 
vanishes at $t_s$ and therefore, the volume of the spacelike 
hypersurface of simultaneity $t = t_s$ is zero. To determine the nature 
of the {\em "anomalous"} point $t_s$, we consider the curvature invariants 
${\cal R}_{\mu \nu} {\cal R}^{\mu \nu}$ and ${\cal R}_{\mu \nu \kp \lm} 
{\cal R}^{\mu \nu \kp \lm}$ (Ryan and Shepley 1975). For $t = t_s$ we obtain 
\be
{\cal R}_{\mu \nu} {\cal R}^{\mu \nu} = \infty = 
{\cal R}_{\mu \nu \kp \lm} {\cal R}^{\mu \nu \kp \lm} 
\ee
and hence the associated hypersurface corresponds to an {\em essential physical 
singularity} (D' Inverno 1993). In contrast to the initial one, at $t = 0$, this 
singularity exists in the future of the spacetime background. 
Nevertheless, at $t = t_s$, the external Hubble parameter remains finite 
and the corresponding scale function differs from zero. 

\section*{3. Renormalizability conditions}

\subsection*{3.1. The effective action method}

The geometrical features of the model (6) are relevant to particle creation 
due to the cosmological evolution of the extra dimension. Nevertheless, 
before computing the macroscopic quantities arising from quantum particle 
production at the one-loop approximation (such as the expectation value of 
the energy-momentum tensor $< T_{\mu \nu} > = < 0_{out} \vert T_{\mu \nu} 
\vert 0_{in} >$ for the created matter), we need, by a general method, to 
establish whether these quantities are well-defined and finite (or at least 
renormalizable) or not. To do so, we consider the {\em effective action} 
method within the context of the path-integral quantization procedure in a 
$C^{\infty}$, $n$-dimensional, globally hyperbolic spacetime. The 
differentiability conditions ensure the existence of differential equations 
and the global hyperbolicity ensures the existence of Cauchy hypersurfaces.

One way to define the {\em effective action} $(W)$ in a curved spacetime, 
is by using the identity
\be
{2 \over \sqrt {-g}} {\dl W \over \dl g^{\mu \nu}} = {< 0_{out} \vert 
T_{\mu \nu} \vert 0_{in}> \over <0_{out} \vert 0_{in}>}
\ee
which is the quantum analogue of the classical variational principle, at 
the one-loop approximation (Birrell and Davies 1982). In this case, one needs 
only to examine whether $W$ is finite or not, since the divergences in $W$ 
are, of course, the same as those that afflict $<T_{\mu \nu}>$. In what 
follows, we treat the number of the spacetime dimensions $(n)$ as a 
variable which can be analytically continued throughout the complex plane 
(dimensional regularization).

The central quantity in the effective action method is the {\em effective 
action generating functional} in the absence of external source currents 
(free field in a curved spacetime). This gives the {\em vacuum persistence 
amplitude} in a curved background (Fischetti et al. 1979, Hartle and Hu 1979, 
Hartle and Hu 1980, Birrell and Davies 1982)
\be
Z[0] = <0_{out} \vert 0_{in}> = \int {\cal D}[\Phi] \; e^{ i \left [ S_G 
(g_{\mu \nu}) + S_m (\Phi) \right ] }
\ee
In Eq. (12), $S_G(g_{\mu \nu})$ is the gravitational action of the 
background field $g_{\mu \nu} (x)$, $S_m(\Phi)$ is the matter action of the 
quantum field and ${\cal D}[\Phi]$ is an appropriate functional measure. In 
the simple theory we are considering, the matter action is no more than 
quadratic in the quantum field.

In computing the functional integral (12) we may set the background field
equal to a solution of the classical field equations, since the error we make 
in doing so is of two-loop order (Jackiw 1974). Then Eq. (12) is written in 
the form
\be
Z[0] = e^{i S_G} \; \int {\cal D}[\Phi] \; e^{- {1 \over 2} i \int d^n y 
\sqrt {-g(y)} \int d^n x \sqrt {-g(x)} \Phi_x K_{x y} \Phi_y}
\ee
In Eq. (13), $K_{x y}$ denotes the differential operator
\be
K_{x y} = \left [ \Box_x + m^2 - i \eps + \ks {\cal R} \right ] 
\dl^n (x - y) {1 \over \sqrt {-g(y)}}
\ee
where $m$ is the mass of the field's quanta and the infinitesimal factor 
$i \eps$ (which is related to Feynman boundary 
conditions on $\Phi$) is used to make the functional integral convergent. 
Gaussian functional integration (Ryder 1985) of Eq. (13), yields
\be
Z[0] = (2 \pi i)^{n/2} \; e^{ i S_G(g_{\mu \nu})} \; \left [ det \: K^{1/2} 
\right ]^{-1}
\ee
In what follows we omit the numerical factor $(2 \pi i)^{n/2}$, since it is 
constant and it may be absorbed into the measure, a thing that simply amounts 
to a redefinition of ${\cal D} [\Phi]$. In any case, that term is 
metric-independent and therefore irrelevant to particle production.

The effective action, $W(\Phi)$, is defined by (Fischetti et al. 1979, Hartle 
and Hu 1979, Hartle and Hu 1980, Birrell and Davies 1982)
\be
Z[0] = <0_{out} \vert 0_{in}> = e^{i W(\Phi)} \; \Rarrow \; 
W(\Phi) = - i \ln {<0_{out} \vert 0_{in}>}
\ee
Then, from Eq. (15) we obtain
\be
W(\Phi) = S_G(g_{\mu \nu}) + {i \over 2} \: tr \: \ln {K}
\ee
The fist term appearing in Eq. (17) is the classical gravitational action. 
The second term represents the one-loop quantum correction to the classical 
action and is relevant to particle production. Indeed, according to the 
defintion of {\em particle production probability} (Itzykson and Zuber 1985), 
from Eq. (16) we find that the total probability to produce a particle 
pair over the entire history of the Universe $(P)$, is given by the formal 
expression
\be
P = 2 \: Im \: W
\ee
However, in Eq. (17) the action term $S_G(g_{\mu \nu})$ must be real, since 
we have considered only classical solutions of the equation that governs the 
propagation of the gravitational field (Hartle and Hu 1980). Therefore, we 
obtain
\be
P = i \: tr \: \ln {K}
\ee
where the operator $K$ is taken to act on a space of vectors, $\vert x >$, 
normalized by
\be
<x \vert x^{\prime}> = \dl^n (x - x^{\prime}) {1 \over \sqrt {-g(x)}}
\ee
According to Eq. (17), if there are any divergences included in the effective 
action (and therefore in $<0_{out} \vert T_{\mu \nu} \vert 0_{in}>$ also), 
those are contained into $P$. Therefore, it is essential to examine under 
which conditions this quantity may become either finite or renormalizable, 
in the one-loop approximation. 

\subsection*{3.2. Renormalizability criteria}

Variation of both sides of Eq. (19) leads to 
\be
\dl P = i \: tr \: \left [ K^{-1} \dl K \right ]
\ee
Now, the identity
\be
K^{-1} = i \: \int_0^{\infty} ds \: e^{-i K s}
\ee
can be used, to obtain
\be
\dl P = - tr \: \int_0^{\infty} ds \: e^{-i K s} \: \dl K
\ee
which, since both $\dl$ and $K$ are independent of $s$, leads to
\be
\dl P = - i \: \dl \left [ tr \: \int_0^{\infty} \: {ds \over s} \: 
e^{- i K s} \right ]
\ee
Accordingly, Eq. (24) leads to the definition
\be
P = -i \: tr \: \int_0^{\infty} \: {ds \over s} \: e^{- i K s}
\ee
up to an arbitrary integration constant, which in any case may be fixed 
by renormalization and therefore can be ignored. The trace of the operator 
$e^{-i K s}$, which acts on the space of vectors given by Eq. (20), is 
defined by (Birrell and Davies 1982)
\be
tr \: e^{- i K s} = \int d^n x \sqrt {-g} \: <x \vert e^{-iKs} \vert x>
\ee
Now, in order to make sense of the formal expression (25) [or (19)] we 
need a representation for the operator $e^{-iKs}$. To do so, we recall 
that the combination of Eq. (14) with the normalization condition
\be
\int \: d^n y \: {1 \over \sqrt {-g(y)}} \: K_{xy} \: K_{y z}^{-1} = 
\dl^n (x - z) \: {1 \over \sqrt {-g(z)}}
\ee
and the equation for the Feynman propagator $G_F(x,x^{\prime})$ in a 
curved spacetime
\be
\left [ \Box_x + m^2 - i \eps + \ks {\cal R} \right ] G_F(x,x^{\prime}) 
= - {1 \over \sqrt {-g(x)}}d^n (x - x^{\prime})
\ee
(Birrell and Davies 1982) lead to
\be
K_{x,x^{\prime}}^{-1} = G_F(x,x^{\prime})
\ee
In this case, it is the existence of the $i \eps$ 
factor into Eq. (28), which ensures that $G_F(x,x^{\prime})$ represents the 
expectation value, in some set of states, of a time-ordered product of fields
\be
G_F(x,x^{\prime}) = <x \vert G_F \vert x^{\prime}>
\ee
If we, furthermore, use Eqs (29) and (30) together with the identity (22), 
we obtain
\be
G_F(x,x^{\prime}) = -i \: \int_0^{\infty} ds \: <x \vert e^{-iKs} \vert 
x^{\prime}>
\ee
For small geodesic distances between two events, the Feynman Green function 
$G_F(x,x^{\prime})$ in curved spacetime, has a nice asymptotic representation, 
called the De Witt-Schwinger representation (De Witt 1975, Bunch and Parker 
1979) 
\be
G_F(x,x^{\prime}) = {1 \over (4 \pi )^{n/2}} \: \Dl^{1/2}(x,x^{\prime} \; 
\int_0^{\infty} \: (is)^{-n/2} \: e^{- i m^2 s + {\sg \over 2is}} \; 
{\cal F} (x,x^{\prime}; is) \; ds
\ee
where $\sg = {1 \over 2} \ks_{\al} \ks^{\al} s^2$ is half of the square of 
the proper distance between $x$ and $x^{\prime}$ ($\ks^{\al}$ is the unit 
vector tangent to the geodesic between $x$ and $x^{\prime}$, at the point 
$x^{\prime}$) and $\Dl (x,x^{\prime})$ is the Van Vleck determinant, defined 
by
\be
\Dl (x,x^{\prime}) = - det \: \left [\partial_{\mu} \partial_{\nu} \sg 
(x,x^{\prime}) \right ] {1 \over \sqrt {g(x) g(x^{\prime})}}
\ee
Notice that, in normal coordinates $(y^{\mu} = \ks^{\mu} s)$ around 
$x^{\prime}$, $\Dl$ reduces to $[g(x)]^{-1/2}$. The function ${\cal F} 
(x,x^{\prime};is)$ is given by the asymptotic expansion (De Witt 1975)
\be
{\cal F} (x,x^{\prime};is) = \sum_{j=0}^{\infty} \: \al_j (x,x^{\prime}) \: 
(is)^j
\ee
with
\bea
\al_0 (x,x^{\prime}) & = & 1 \nn \\
\al_1 (x,x^{\prime}) & = & ({1 \over 6} - \ks ) {\cal R} - {1 \over 2} 
({1 \over 6} - \ks) {\cal R}_{; \mu}y^{\mu} - {1 \over 3} \al_{\mu \nu} 
y^{\mu} y^{\nu} \nn \\
\al_2 (x,x^{\prime}) & = & {1 \over 2} ({1 \over 6} - \ks)^2 {\cal R}^2 
+ {1 \over 3} \al_{\lm}^{\lm}
\eea
where
\bea
\al_{\mu \nu} & = & {1 \over 2} (\ks - {1 \over 6}) {\cal R}_{; \mu \nu} 
+ {1 \over 120} {\cal R}_{; \mu \nu} - {1 \over 40} {\cal R}_{\mu \nu ; 
\lm}^{\lm} - {1 \over 30} {\cal R}_{\mu}^{\lm} {\cal R}_{\lm \nu} \nn \\
& . & \nn \\ 
& + & {1 \over 60} {\cal R}^{\kp \lm}_{\mu \nu} {\cal R}_{\kp \lm} 
+ {1 \over 60} {\cal R}^{\lm \rh \kp}_{\mu} {\cal R}_{\lm \rh \kp \nu}
\eea
and the rest of the $\al_j$'s (for $j > 2$) are given by certain recursion 
relations (De Witt 1975, Christensen 1976). Now, using Eqs. (31) and (32) 
we obtain
\be
<x \vert e^{-i K s} \vert x^{\prime}> = {i \over (4 \pi )^{n/2}} \: 
\Dl^{1/2}(x,x^{\prime}) \: (is)^{-n/2} \: e^{- i m^2 s + {\sg \over 2is}} 
\: {\cal F} (x,x^{\prime}; is) 
\ee
and the combination of Eqs. (25), (26) and (37) leads to
\be
P = \lim_{x \rarrow x^{\prime}} \; \int \: d^n x \sqrt {-g} {1 \over 
(4 \pi )^{n/2}} \: \Dl^{1/2}(x,x^{\prime}) \: \int_0^{\infty} \: (is)^{-n/2} 
\: e^{- i m^2 s + {\sg \over 2is}} \: {\cal F} (x,x^{\prime}; is) \: 
{ds \over s}
\ee
If the number of the spacetime dimensions can be treated as a variable 
which can be analytically continued throughout the complex plane, then we 
may take the $ x \rarrow x^{\prime}$ limit, to obtain
\be
P = i \int \: d^n x \sqrt {-g} {1 \over (4 \pi )^{n/2}} \: \sum_{j=0}^{\infty} 
\al_j (x) \: \int_0^{\infty} \: (is)^{j- 1 - {n \over 2}} \: e^{- i m^2 s} \: 
ds
\ee
which yields 
\be
P = {1 \over (4 \pi)^{n/2}} \: \sum_{j=0}^{\infty} \: m^{n - 2j} \: 
\Gm (j - {n \over 2}) \; \int \: d^n x \sqrt {-g} \: \al_j (x)
\ee
where, now, the quantities $\al_j$ become
\be
\al_j (x) = \al_j (x,x^{\prime}) \vert_{s=0} \sim {\cal R}^j (x)
\ee
i.e. polynomials in scalar combinations of the Riemann curvature tensor 
and its contractions. Therefore, the particle production probability is 
manifestly positive and covariant under real coordinate transformations. 
Nevertheless, from Eq. (40) it is evident that P may diverge, because of 
poles in the $\Gm$-functions. Notice, for instance, that for $n = 4$ the 
first three terms in Eq. (40) are divergent.

The question that arises now is, whether there exists some special fixing 
of the parameters involved, in order to make the particle production 
probability finite for every value of the index $j$ or at least, as far 
as we are concerned, up to $j=2$ i.e. for a gravity theory quadratic in 
the curvature 
tensor. We conclude that the simplest fixing of this sort is that the 
spacetime dimension is an {\bf odd} number. Then, the r.h.s. of Eq. (40) is 
not only renormalizable, but actually {\bf finite}. According to Eqs. (17) 
and (11), this is also true for both the vacuum persistence amplitude and 
the vacuum expectation value of the energy-momentum tensor for a scalar 
field in a $C^{\infty}$, globally hyperbolic, $n$-dimensional 
spacetime. Since in a curved background $\vert 0_{out}> \neq \vert 0_{in}>$ 
even in the absence of external current sources, we are forced to consider an 
{\em odd-dimensional} spacetime in order to deal with a finite, well-defined 
probability of particle production, without using any renormalization technique. 
The most simple extrapolation of the four-dimensional spacetime to an 
odd-dimensional one, is a curved background of {\bf five dimensions}.

Notice that, for $n$ odd, in Eq. (40) we have $P \neq 0$ only as long as 
$m \neq 0$ and hence, no 
massless particles are actually produced. This result is probably a 
mathematical template, related to the short-distance approximation used in 
the derivation of of the representation (32) for Feynman's Green function 
in curved spacetime (De Witt 1975, Christensen 1976, Bunch and Parker 1979). 
Nevertheless, short distances are probed by the high-frequency modes of a 
quantum field (Birrell and Davies 1982) and a high-frequency approximation may 
become quite accurate as regards particle production in the neighbourhood of 
the singular epoch $t = t_s$.

Indeed, when a particle is created at a proper distance $d_s$ from a 
singularity, it 
should {\em "fit"} inside the corresponding area (Nesteruk 1991) and 
therefore its wavelength should satisfy the condition $\lm \leq d_s$. 
Accordingly, for the metric (6) we obtain
\be
\om^2 \geq 4 \pi^2 \: {t_s^4 \over \left ( {t \over t_s} -1 \right )^2}
\ee
In the immediate vicinity of the singularity, that is when $t \rarrow t_s$, 
Eq. (42) results to $\om^2 \gg 1$ and therefore, we are led to examine only 
the high-frequency behaviour of the quantum field under consideration. It 
is therefore probable that no massless scalar quanta are created in the 
neighbourhood of the singular epoch $t = t_s$.

On the other hand, the fact that no particles are produced in the external 
space of the metric (6) when $m = 0$, is in complete agreement with a 
corresponding result obtained using thermodynamical considerations (Kleidis 
and Papadopoulos 1997a). Indeed, the metric (6) is an exact solution of 
the classical gravitational field equations when the pressure of the 
internal (one-dimensional) space equals to zero $(p_{int} = 0)$. According to 
Kleidis and Papadopoulos (1997a), as regards the thermodynamical description 
of two subspaces (the external space and the internal one), this condition 
corresponds to an {\em adiabaticity criterion} and therefore, 
no radiation (massless particles) is allowed to be transfered from one 
space to the other. Both subspaces represent {\em closed} thermodynamical 
systems (see also Prigogine et al 1989). However, when $m \neq 0$, we do not 
have exchange of energy (radiation) between the two subspaces, but exchange 
of matter (particles) where the total number of particles in each space is no 
longer constant ($\dl N \neq 0$). Then, the problem is reduced to that of the 
{\em open} thermodynamical systems, which allows for particle production in 
the visible space (Prigogine 1961, Prigogine et al. 1989, Kleidis and 
Papadopoulos 1997a). 

In what follows, we solve explicitly the resulting Klein-Gordon wave equation, 
to obtain exact mode solutions for the scalar field in the five-dimensional 
spacetime under consideration, in order to study massive particle creation 
in four dimensions due to the cosmological evolution of the fifth dimension. 
It has been recently shown that the four-dimensional external space of the 
model (6) is stable against perturbations in the energy density (Kleidis et al. 
1996) and hence its dynamics will be driven by other factors than just particle 
production. Accordingly, we shall ignore {\em backreaction} of the created 
particles on the external background.

\section*{4. Particle creation}

\subsection*{4.1. Quantum field in curved spacetime}

At the one-loop level, the (semiclassical) interaction between a massive 
quantum scalar field $\Ph (t, \vec{x})$ and a classical gravitational one 
(the five-dimensional spacetime under consideration), may be determined 
through Hamilton's principle, involving the action scalar (Calan et al. 1970, 
Birrell and Davies 1982)
\be
S_m (\Ph) = {1 \over 2} \int \sqrt {-g(x)} \left [ g^{\mu \nu} (x) 
\Ph_{, \mu}(x) \Ph_{, \nu}(x) - \left ( m^2 + \ks {\cal R}(x) \right ) 
\Ph^2(x) \right ] d^5 x
\ee
where $m$ is the five-dimensional mass of the field's quanta (Kleidis and 
Papadopoulos 1997a) and $\ks$ is a dimensionless coupling constant. Euler 
variation of Eq. (43) leads to the wave equation that governs the 
propagation of the massive scalar field in a five-dimensional spacetime 
(Friedlander 1975).

The quantization of the field $\Ph (t, \vec{x})$ is performed by 
imposing canonical commutation relations on the hypersurface 
$t =$ constant (Isham 1975, 1981) 
$$
\left [ \Ph (t, \vec{x}) \; , \; \Ph (t, {\vec{x}}^{\prime}) \right ] 
= \left [ \pi (t, \vec{x}) \; , \; \pi (t, {\vec{x}}^{\prime}) 
\right ] = 0
$$
\be
\left [ \Ph (t, \vec{x}) \; , \; \pi (t, {\vec{x}}^{\prime}) 
\right ] = i \dl^{(3)} \left ( \vec{x} - {\vec{x}}^{\prime} \right ) 
\ee
where $\pi (t, \vec{x})$ is the momentum canonically conjugated to the 
field $\Ph(t, \vec{x})$. Once an appropriate definition of the {\em "in"} 
and the {\em "out"} vacuum states is given (Fulling 1979), the quantum field 
operator, $\Ph (t, \vec{x})$, may be expanded in terms of creation $(A_{k,k_5})$ 
and annihilation $(A_{k,k_5}^{\dagger})$ operators (Birrell and Davies 1982, 
Garriga and Verdaguer 1989), as follows
\be
\Ph (t, \vec{x}) = {1 \over 4 \pi^2} \sum_{k_5 = 0}^{\infty} 
\int d \mu (k) \left [ A_{k,k_5} u_{k,k_5} (t, \vec{x}) + A_{k,k_5}^{\dagger} 
u_{k,k_5}^{\ast} (t, \vec{x}) \right ]
\ee
where $u_{k,k_5}$ is a complete, orthonormal set of field modes (the solutions 
of the equation of propagation) and $d \mu (k)$ an appropriate measure depending 
on the FRW model considered for the external space (Grib et al. 1976, 1980). 
In Eq. (45) $k, k_5$ are quantum numbers labeling a particular mode. The 
resulting spectrum is continuous in $k$ $(0 \leq k < \infty)$ and discrete in 
$k_5$. In fact, since the internal space is assumed compact, we have 
\be
k_5 = {\ell \over R_5}
\ee
where $R_5$ is the {\em "radius"} of the internal space and $\ell$ an integer.
In what follows, we focus attention on modes with $k_5 \neq 0$ which may be 
associated to massive particles in the four-dimensional sense (e.g. see 
Garriga and Verdaguer 1989).

To interprete the quantized field $\Ph(t, \vec{x})$ in terms of particles, we 
need to give an appropriate definition of the positive frequency modes, 
$u_{k,k_5}(x)$, in the {\em "in"} and the {\em "out"} vacuum states (Birrell 
and Davies 1982). For most cosmological models it is difficult to find a 
natural candidate for the {\em in-modes} (Isham 1981), due to the ill-defined 
nature of the particle concept in a curved spacetime (Wald 1984). This 
difficulty is usually circumvented by assuming that before some initial time 
$t_{in}$, larger than the Planck time $t_{Pl}$, the spacetime is matched to 
be static (Hu 1974). This is justified by the fact that at the Planck epoch, 
the semiclassical approach which was taken into account, must break down. 
Therefore, on mathematical grounds, in order to discuss particle creation 
we need to exclude the influence of the period $ 0 < t < t_{Pl} \sim 10^{-43}$ 
sec and to give an unambiguous definition for the positive frequency modes 
$u_{k,k_5}^{in}$. Hence, it seems reasonable to assume a static 
{\em in-metric}. Then, the in-vacuum is equivalent to the Minkowski vacuum 
(Candelas and Dowker 1979, Birrell and Davies 1982). As regards the {\em 
out-modes} we note that, as $t \rarrow \infty$ (out-region), the internal 
scale function becomes static, $S^2 (t) \rarrow 1$, while the external one 
varies at a very low rate
\be
{d^{\ell} \over dt^{\ell}} \left ( {{\dot R} \over R} \right ) 
\rarrow 0 \; \; , \; \; \; \ell \geq 0 
\ee
and therefore, the visible space evolves adiabatically (Fulling et al 1974, 
Davies and Fulling 1977, Bunch et al 1978). Then, the {\em out-vacuum} state 
corresponds to an {\em "adiabatic vacuum"}, where the positive frequency modes 
are well-defined and no particles are created (see also Birrell and Davies 
1982, Garriga and Verdaguer 1989).

Those two vacuum states are related by means of a Bogolubov transformation 
(Bogolubov 1958)
\be
u_{k,k_5}^{in} = \al_{k,k_5}(t) u_{k,k_5}^{out} + \bt_{k,k_5}(t) 
{u_{-k,-k_5}^{out}}^{\ast}
\ee
where the functions $\al_{k,k_5}(t)$ and $\bt_{k,k_5}(t)$ are the Bogolubov 
coefficients. Initially (in the {\em in-region}), we have $\al_{k,k_5}(t_{in}) 
= 1$ and $\bt_{k,k_5}(t_{in}) = 0$ (Zel'dovich and Starobinsky 1972, Hu and 
Parker 1978). As $t \rarrow \infty$ (in the {\em out-region}), we expect that 
these functions reduce to constant values, different from those at $t = t_{in}$, 
which give us the linear combination of positive and negative frequency 
out-modes that makes up a positive frequency in-mode. Then, the number of 
particles created in the mode $(k,k_5)$ is given by (Birrell and Davies 1982)
\be
N_{k,k_5} = \vert \bt_{k,k_5} \vert^2
\ee
Therefore, in order to calculate the number of particles created in the 
spacetime (6), we need to determine the Bogolubov coefficient $\bt_{k,k_5}(t)$ 
at $t = \infty$. By virtue of Eq. (48), this may be performed by solving the 
equation of propagation for the quantum scalar field in the curved spacetime 
under consideration.

Minimizing the variation of the action (43) with respect to $\Ph (t, \vec{x})$, 
we are led to the following equation of propagation for the quantum field 
\be
\Box \: u_{k,k_5}(x) + \left [ m^2 + \ks {\cal R}(x) \right ] \: u_{k,k_5}(x) 
= 0
\ee
To separate the variables in Eq. (50), we substitute
\be
u_{k,k_5} (\ta, \vec{x}, x^5) = {1 \over R(\ta) S^{1/2}(\ta)} \: 
{\cal Y}_k (\vec{x}) {\ch}_{k,k_5}(\ta) \: e^{i k_5 x^5}
\ee
where $ \vec{x} = (r, \th , \ph)$ and
\be
\ta = \int_{t_s}^t {dt \over R(t)} = ln { \left ( {t \over t_s} 
\right ) }
\ee
is the {\em conformal time}, with its origin placed at $t = t_s$. In Eq. (51), 
the quantities ${\cal Y}_k (\vec{x})$ are the eigenfunctions of the operator 
$\triangle^{(3)}$
\be
\triangle^{(3)} \: {\cal Y}_k (\vec{x}) = - (k^2 + 1) \: {\cal Y}_k (\vec{x})
\ee
which corresponds to the Laplacian associated with the hyperbolic spatial 
metric of the external space, in the model (6). The harmonics ${\cal Y}_k 
(\vec{x})$ are normalized according to
\be
\int \: d^3 x \: \sqrt {^{(3)}g} \: {\cal Y}_k (\vec{x}) \: 
{\cal Y}_{k^{\prime}}^{\ast} (\vec{x}) = \dl (k,k^{\prime})
\ee
where $\sqrt {^{(3)}g}$ is the determinant associated to the three-dimensional 
external metric and $\dl (k,k^{\prime})$ is the $\dl$-function with respect to 
the measure $\mu (k)$
\be
\int \: d \mu (k) \: f(k^{\prime}) \: \dl (k,k^{\prime}) = f(k)
\ee
In this case, the eigenfunctions of the three-dimensional Laplacian are given 
by (Parker and Fulling 1974)
\be
{\cal Y}_k (\vec{x}) = \Pi_{k J}^{(-)} (r) \: Y_J^M (\th, \ph)
\ee
where $k = (k, J, M)$ and
$$
0 \leq k < \infty
$$
\bea
J & = & \; \; \; 0, \; 1, \; 2, ... \nn \\
M & = & -J ... 0 ... J
\eea
In Eq. (56), the functions $Y_J^M (\th, \ph)$ are spherical harmonics. As 
regards the functions $\Pi_{k J}^{(-)} (r)$, they are defined by (Lifshitz 
and Khalatnikov 1963, Bander and Itzykson 1966, Ford 1976)
\be
\Pi_{k J}^{(-)} (r) = \lbrace {1 \over 2} \pi k^2 (k+1)^2 ... 
[k^2 + (2+J)^2] \rbrace^{-1/2} \: \sinh {\ch} \: \left 
( {d \over d \cosh {\ch}} \right )^{1+J} \cosh {k \ch}
\ee
where $\ch = {\sinh {r}}^{-1}$. Finally, the measure $\mu (k)$ is defined as 
follows
\be
\int \: d \mu (k) = \int_0^{\infty} \: \sum_{J,M}
\ee

\subsection*{4.2. Particle creation and the mass-energy spectrum}

Taking into account the above definitions and introducing Eq. (51) into (50), 
we obtain the following equation for the time-dependent part of the modes 
$\ch_{k,k_5} (\ta)$
\be
\ch_{k,k_5}^{\prime \prime} (\ta) + \left [ k^2 + R^2 (\ta) 
\left ( {k_5^2 \over S^2 (\ta)} - m^2 \right ) + {1 \over 
(e^{2 \ta} - 1)^2} \right ] \: \ch_{k,k_5} (\ta) = 0
\ee
where $R^2 (\ta)$ and $S^2 (\ta)$ are the scale functions of the external and 
the internal space respectively, in terms of the conformal time and a 
prime denotes differentiation with respect to $\ta$. Notice that, since the 
metric (6) is Ricci flat, the propagation of the scalar field in the curved 
spacetime under consideration is no longer dependent on the value of the 
coupling constant $\ks$. The orthonormality of the modes $u_{k,k_5}(x)$, 
according to the scalar product in curved spacetime (Birrell and Davies 1982), 
in connection to normalization conditions (54) and (55), result to the 
following Wronskian relation on $\ch_{k,k_5}(\ta)$
\be
{\ch_{k,k_5}^{\ast}}^{\prime} \: \ch_{k,k_5} - \ch_{k,k_5}^{\prime} \: 
\ch_{k,k_5}^{\ast} = i
\ee
Eq. (60) is essentially five-dimensional. In order to study massive particle 
creation in the visible space of the metric (6), at first, we have to determine 
the four-dimensional analogue of the five-dimensional mass for the field's 
quanta. It has been recently suggested that, in the context of a 
$1+3+D$-dimensional spacetime (where a D-dimensional, compact internal space 
is present), the matter-energy content of a volume element in three dimensions 
is given by (Kleidis and Papadopoulos 1997a)
\be
{\cal E}_3 = {\cal E}_{3+D} \: S^D
\ee
where ${\cal E}_{3+D}$ is the corresponding matter-energy content in $3+D$ 
dimensions and $S^D$ is the proper volume of the (closed and bounded) 
internal space. In this case, the overall matter-energy content of 
a $3+D$-dimensional volume element is being projected onto its 
three-dimensional counterpart (see also Kolb et al. 1984, Farina-Busto 1988), 
which is in complete agreement with the definition of the {\em stress-energy 
tensor on a hypersurface} (Misner et al. 1973). In the absence of the extra 
dimensions $(D = 0)$ we obtain 
${\cal E}_3 = {\cal E}_{3+D}$, corresponding to the matter-energy content of 
a volume element in a four-dimensional spacetime. Accordingly, we consider the 
volume element occupied by a subatomic particle in five-dimensions $(D = 1)$ 
and its four-dimensional projection. Then, Eq. (62) leads to the following 
relation between the mass-energy of a particle in a five-dimensional spacetime 
$(m)$ and its four-dimensional counterpart $(m_4)$
\be
m_4 = m \: S
\ee
In this case, $m_4$ may be identified as the {\em phenomenological mass} of a 
particle in four dimensions, when a compact, one-dimensional, time-dependent 
internal space is present (in this respect, see also Wesson 1992, Mashhoon et 
al 1994). Now, inserting Eq. (63) into Eq. (60), we obtain
\be
\ch_{k,k_5}^{\prime \prime} (\ta) + \left [ k^2 + {R^2 \over S^2 } 
\: \left ( k_5^2 - m_4^2 \right ) + {1 \over (e^{2 \ta} - 1)^2} \right ] \: \ch_{k,k_5} 
(\ta) = 0
\ee
The solutions of the wave equation (64) describe the exitation of 
four-dimensional, 
massive scalar modes, which takes place as a result of the large scale 
anisotropic cosmological evolution between the internal and the external 
space. This is due to the fact that these modes 
possess an additional, non-vanishing momentum component along the fifth 
dimension $(k_5 \neq 0)$. Then, at times $- \infty < \ta < 0$ where the 
internal space contracts at high rate, the {\em physical frequency} of 
the scalar field along the extra dimension
\be
\om_5 (k_5) = {k_5 \over S (\ta)}
\ee
is blueshifted, to bring about particle creation at high energies, in the 
neighbourhood of the singular epoch $ \ta = 0 \; (t = t_s)$ [e.g. see 
Eq. (42)]. The energies of the created particles are subsequently decreased 
in the region $0 < \ta < + \infty$, where the internal space expands at a 
low rate (Fig. 1) and the fifth-dimensional frequency component of the modes 
is accordingly redshifted.

Eq. (64) is reminiscent of the classical equation of motion for a 
harmonic oscillator, where the coefficient of the linear term $\ch_{k,k_5} 
(\ta)$ may be considered as representing a time-dependent frequency parameter. 
At $\ta \rarrow - \infty$ (the in-region), as well as at $\ta \rarrow + 
\infty$ (the out-region), this parameter must settle down to non-negative 
constant values, since no mode excitation takes place in these regions. 
As $\ta \rarrow \pm \infty$, the modes of the quantum field are the 
well-defined positive frequency modes and no particles are created (Birrell 
and Davies 1982). Accordingly, in the search for solutions of Eq. (64) 
that exhibit a 
nice-behaviour in both the "in" and the "out" vacuum states, we demand
\be
k_5 - m_4 = 0
\ee
or in physical units
\be
m_4 c^2 = \hbar c k_5
\ee
which, by virtue of Eq. (46), is written in the form
\be
\; \; \; m_4 c^2 = \ell \: {\hbar c \over R_5} \; , \; \; \ell = 1, 2, ...
\ee
thus, resulting in
\be
m_4 c^2 = \ell \: \times \: 3.1638 \times 10^{-17} {1 \over R_5} \; \; (erg)
\ee
or
\be
m_4 c^2 = \ell \: \times \: 1.977375 \times 10^{-11} {1 \over R_5} \; \; (MeV)
\ee
where, in both cases, $R_5$ is measured in $cm$.

Eq. (68) represents the energy spectrum attributed to the scalar field in 
four dimensions as a 
consequence of the cosmological evolution of the five-dimensional spacetime 
under consideration. We see that, because of condition (66), two major 
aspects of the theory are revealed: {\bf (i)} The spectrum is discrete (and 
therefore the energy is quantized) as a result of compactification of the 
extra dimension. {\bf (ii)} The mass-energy of the particles produced (the 
quanta of the scalar field) is related to the inverse radius of the internal 
space and therefore it is crucially dependent on the exact value of $R_5$.

By virtue of condition (66), the equation which describes the creation of 
massive particles in the visible space of the spacetime (6), is written in 
the form
\be
\ch_{k,k_5}^{\prime \prime} (\ta) + \left [ k^2 + {1 \over (e^{2 \ta} - 1)^2} 
\right ] \: \ch_{k,k_5} (\ta) = 0
\ee
Notice that the corresponding analysis is no longer dependent on the 
four-dimensional mass 
parameter $(m_4)$. Eq. (71) may be written in a more convenient form, as 
follows
\be
\ch_{k,k_5}^{\prime \prime} (\ta) + \left [ k^2 + {1 \over 4} \left ( 
1 - \coth {\ta} \right )^2 \right ] \ch_{k,k_5} (\ta) = 0
\ee
To solve Eq. (72), we perform a substitution in the dependent variable, 
of the form 
\be
\ch_{k,k_5}(\ta) = e^{-i \Om(\ta)} F(\ta)
\ee
where
\be
\Om(\ta) = \om_+ \ta + \om_- \ln {(\sinh {\ta})}
\ee
and 
\be
\om_+ = {1 \over 2} \left ( \sqrt {k^2 + 1} + k \right )
\ee
\be
\om_- = {1 \over 2} \left ( \sqrt {k^2 + 1} - k \right )
\ee
together with the following substitution in the independent variable
\be
z = {1 \over 2} \left ( 1 + \coth {\ta} \right )
\ee
Then, Eq. (72) is reduced to a hypergeometric equation for $F(z)$ of the 
form
\be
z (1 - z) {d^2 F \over d z^2} + \left [ c - (a + b + 1) z \right ] 
{d F \over d z} - a b F = 0
\ee
where
\be
a = {1 \over 2} + i \om_- = b
\ee
and 
\be
c = 1 - i k
\ee
This equation can be solved analytically. Since the function $\coth {\ta}$ 
consists of two irrelative branches (one for $\ta < 0$ and the other for 
$\ta > 0$), the corresponding analysis is accordingly separated in two cases, 
one for each sign of $\ta$. Then, the time-dependent part 
of the normalized modes which behave like the positive frequency Minkowski 
modes in the remote past ($\ta < 0$ and $\ta \rarrow - \infty$), is
\bea
\ch_{k,k_5}^{in}(\ta) & = & {1 \over \sqrt {2 \om_<}} 
e^{- i \left [ \om_+ \ta + \ln {(\sinh {\ta})} \right ]} \nn \\
& \times & _2 F_1 \left ( {1 \over 2} + i \om_- \; , \; 
{1 \over 2} + i \om_- \; ; \; 1 - i \om_< \; ; \; 
{1 \over 2} \left [1 + \coth {\ta} \right ] \right )
\eea
where $\om_< = k$. On the other hand, the time-dependent part of the modes which 
behave like positive frequency modes in the remote future ($\ta > 0$ and $\ta 
\rarrow \infty$), is found to be
\bea
\ch_{k,k_5}^{out}(\ta) & = & {1 \over \sqrt {2 \om_>}} 
e^{- i \left [ \om_+ \ta + \ln {(\sinh {\ta})} \right ]} \nn \\
& \times & _2F_1 \left ( {1 \over 2} + i \om_- \; , \; 
{1 \over 2} + i \om_- \; ; \; 1 + i \om_> \; ; \; 
{1 \over 2} \left [1 - \coth {\ta} \right ] \right )
\eea
where $\om_> = \sqrt {k^2 + 1}$. Clearly, the solutions (81) and (82) are not 
equal, which means that the Bogolubov coefficient $\bt_{k,k_5}(\ta)$ in Eq. 
(48) is non-vanishing. To determine its value at $\ta \rarrow \infty$ we use 
the linear transformation properties of hypergeometric functions 
(Abramowitz and Stegun 1970). As regards the modes (81) and (82), we verify 
that Eq. (48) holds, provided that
\be
\al_{k,k_5} = \sqrt {\om_> \over \om_<} \; {\Gm \left ( 1 - i \om_< \right ) 
\Gm \left ( -i \om_> \right ) \over \left [ \Gm \left ({1 \over 2} 
- i \om_+ \right ) \right ]^2}
\ee
and
\be
\bt_{k,k_5} = \sqrt {\om_> \over \om_<} \; {\Gm \left ( 1 - i \om_< \right ) 
\Gm \left ( i \om_> \right ) \over \left [ \Gm \left ({1 \over 2} 
+ i \om_- \right ) \right ]^2}
\ee
Furthermore, using the properties of the Gamma function in a complex plane 
(Abramo-witz and Stegun 1970), we find that the number of massive particles 
created in the {\em out-region}, in the mode denoted by $(k, k_5)$, is
\be
N_{k,k_5} = \vert \bt_{k,k_5} \vert^2 = {\cosh^2 { \left (\pi \om_- 
\right )} \over \sinh { \left ( \pi \om_< \right ) } 
\sinh {\left ( \pi \om_> \right) } }
\ee
and that the following normalization condition holds 
\be
\vert \al_k \vert^2 - \vert \bt_k \vert^2 = 1 
\ee
as a consequence of Eq. (61).

Since the quantum state chosen for $\ta \rarrow - \infty$ corresponds to the 
{\em in-vacuum}, in the {\em out-region} the number density per unit proper 
volume and the corresponding energy density (in physical units) of the massive 
particles created in the ordinary space, in a particular mode $k_5$, are (Grib 
et al 1976, 1980, Birrell and Davies 1980, 1982, Anderson and Parker 1987, 
Garriga and Verdaguer 1989, Laciana 1993)
\be
\et = {1 \over 2 \pi^2 c^3 R^3} \int_0^{\infty} k^2 \vert \bt_{k,k_5} \vert^2 dk
\ee
and 
\be
\rh = {\hbar \over 2 \pi^2 c^3 R^4} \int_0^{\infty} k^3 \vert \bt_{k,k_5} 
\vert^2 dk
\ee
where, in our case, $R(t)$ is given by Eq. (7).

Analytic evaluation of the integrals (87) and (88), to estimate the 
corresponding physical quantities, may be carried out only in the ultraviolet 
limit, where $k^2 \gg 1$. According to Eq. (42), this approximation corresponds 
to creation of particles in the immediate neighbourhood of the singularity at 
$t = t_s$ $(\ta = 0)$. In this case, the number of massive particles created 
in the mode $(k, k_5)$, may be written in the form
\be
N_{k,k_5} = \vert \bt_{k,k_5} \vert^2 \simeq {1 \over \sinh^2 {\left ( 
\pi k \right ) }}
\ee
Then, Eq. (87) becomes
\be
\et = {1 \over 2 \pi^2 c^3 R^3} \int_0^{\infty} k^2 
{1 \over \sinh^2 {\left ( \pi k \right )}} dk
\ee
which leads to (Gradshteyn and Ryzhik 1965)
\be
\; \; \; \et = {1 \over 12 \pi^3 c^3 \: t_s^3} \; e^{- 3 \ta} \; \; \; \left (
{particles \over cm^3} \right )
\ee
where $t_s$ is measured in sec. Accordingly, the energy density of the massive 
particles created in ordinary space in the neighbourhood of the singular epoch 
$t = t_s$, is
\be
\rh = {\hbar \over 2 \pi^2 c^3 R^4} \int_0^{\infty} k^3 
{1 \over \sinh^2 {\left ( \pi k \right )}} dk 
\ee
which results to (Gradsteyn and Ryzhik 1965)
\be
\; \; \; \rh = {3 \hbar  \over 4 \pi^6 c^3 \: t_s^4} \; \zt (3) \; 
e^{- 4 \ta} \; \; \; \left ( {erg \over cm^3} \right )
\ee
where $\zt (3)$ is Riemann's Zeta function of argument $3$ $(\zt (3) = 1.202)$. 

We see that, in the out-region (for $\ta \rarrow \infty$) both $\et$ and 
$\rh$ are exponentially suppressed. This is not an unexpected result. The 
production of high-mass particles should be exponentially small, because 
of the large amount of energy which must emerge from the changing gravitational 
field to supply the particles' rest mass. Indeed, for modes with $k_5 \neq 0$, 
the physical frequency in the fifth dimension is strongly blueshifted during 
the contraction branch of the internal space (for $ - \infty < \ta < 0$) and 
the particles produced at $\ta = 0$ aquire energies of the order of Planck 
mass. Consider, for instance, that the radius of the internal space is of 
the order of Planck length, $R_5 = l_{Pl} = 1.616 \times 10^{-33}$ cm, as 
indicated by modern Kaluza-Klein theories (Applequist et al 1987, Kolb and 
Turner 1990). Then, Eq. (70) results in an energy scale for the created 
particles, of the order
\be
m_4 c^2 = \ell \; \times \; 1.223 \times 10^{19} \; \; GeV
\ee
which corresponds to the typical energy scale at the Planck epoch (Planck 
mass). Eq. (94) raises a problem because (when $R_5 = l_{Pl}$), even if a 
single massive mode is excited, its contribution to the energy density 
becomes so large that backreaction should be taken into account and the 
cosmological model (6) that we have started with, breaks down.

Nevertheless, we expect that both the number and the energy density of the 
particles emerging in four dimensions, during the whole creation process 
(for $- \infty < \ta < + \infty$), are subsequently redshifted due to the 
simultaneous cosmological expansion of the visible space. In this context, 
the total number and energy density of the {\em "real"} particles present 
at late times $t$, are given by the sum of all the increments, $\dl \et (t)$ 
and $\dl \rh (t)$ respectively, which were created at earlier times 
$t^{\prime} < t$ and being redshifted by a factor $R(t^{\prime})/R(t)$. 
Therefore, as regards the total number density of the "real" particles 
present at late times, we have (Hu and Parker 1977, 1978, Nesteruk and 
Pritomanov 1990, Nesteruk 1991)
\be
\et_R = \int_{t_{in}}^t \left [ {R(t^{\prime}) \over R(t)} \right ]^3 
\left [ - \: {d  \over d t^{\prime}} \; n(t^{\prime}) \right ] 
d t^{\prime}
\ee
which gives
\be
\et_R (t) = {1 \over 4 \pi^3 c^3 } \; {1 \over t^3} \; \ln {\left ( 
{t \over t_{in}} \right ) }
\ee
while, as regards the corresponding energy density, we have 
\be
\rh_R = \int_{t_{in}}^t \left [ {R(t^{\prime}) \over R(t)} \right ]^4 
\left [ - \: {d  \over d t^{\prime}} \; \rh (t^{\prime}) \right ] 
d t^{\prime}
\ee
from which we obtain
\be
\rh_R (t) = {3 \hbar \over \pi^6 c^3 } \; \zt (3) \; {1 \over t^4} \; 
\ln { \left ( {t \over t_{in}} \right ) }
\ee
Eqs. (96) and (98) represent the quantities which may be identified as the 
{\em "observable"} number and energy density of the created particles in the 
(four-dimensional) out-region. Both $\et_R$ and $\rh_R$ correspond to 
high-frequency modes, representing particles created in the immediate 
neighbourhood of a singular epoch (a narrow time-interval around $t=t_s$). 
Indeed, from Fig. 2 we see that those quantities are highly localized in 
time, around $t = 1.4 t_{in}$, which may be considered as representing the 
singular epoch $t=t_s$. For $t \rarrow t_{in}$, as well as for $t \rarrow 
\infty$, both $\et_R$ and $\rh_R$ vanish and the process of particle production 
ceases, as expected.

Finally, we have to point out that, because of the low rate expansion 
experienced by the internal space for $t > t_s$, the frequency component 
of the modes along the extra dimension $\om_5 (k_5)$, is redshifted in 
the corresponding time-interval. Therefore, we expect that the actual 
values of the observable quantities $\et_R$ and $\rh_R$ will be a 
little bit lower than those predicted by Eqs. (96) and (98).

\section*{5. Discussion and Conclusions} 

In the present article we have considered massive scalar particle production 
from vacuum, as a result of the anisotropic evolution of a five-dimensional 
cosmological model, arising from gravitational theories of quadratic 
Lagrangians. 

The spacetime under consideration is an exact classical solution of the 
cosmological field equations, obtained from a gravity theory which includes 
terms quadratic in curvature tensor (Kleidis and Papadopoulos 1997b), when 
the matter that fills the Universe is in the form of a closed or heterotic 
superstring perfect gas (Matsuo 1987). The manifold consists of two 
anisotropically evolving subspaces (the external space and the internal one) 
and contains two singularities. The first one, at $t = 0$, is essential and is 
excluded from the analysis, together with the Planck epoch for which we have no 
information (Hu 1974, Grib et al 1980, Wald 1984). On the contrary, the second 
singularity, at $t = t_s > 0$, lies in the hart of the particle production 
process. At $t = t_s$ the physical size of the internal space becomes zero, 
while the external Hubble parameter is finite and the corresponding scale 
function non-vanishing. The existence of such {\em internal} singularities may 
affect on compactification of the extra dimension, but it does not affect on 
its {\em "invinsibility"} at the present epoch (McInnes 1985).

In the search for renormalization conditions on the observable quantities 
arising from particle creation, i.e. $<0 \vert T_{\mu \nu} \vert 0>$, we have 
used the effective action method in the one-loop approximation. In terms of 
functional analysis we have found that, in a classical, $C^{\infty}$, 
globally hyperbolic, $n$-dimensional spacetime, the total probability to 
produce a particle pair over the entire history of the Universe is finite, 
provided that the spacetime is odd-dimensional. Since the divergences in 
$P$ are the same as those that afflict both the effective action $W$ and 
the vacuum expectation value of the energy-momentum tensor for the scalar 
field, $<0 \vert T_{\mu \nu} \vert 0>$ (Birrell and Davies 1982), we may 
conclude that the physical, observable quantities relevant to any process of 
particle creation in a curved spacetime are not only one-loop renormalizable 
but actually {\em one-loop finite}, provided that this process is applied on 
a curved background of odd dimensions. The most simple extrapolation of the 
four-dimensional spacetime to an odd-dimensional one is a {\bf five-dimensional 
spacetime}. Indeed, on {\em supersymmetric} considerations (Van Nieuwenhuizen 
1977, Alvarez 1989), one-loop finiteness in five dimensions implies the 
existence of only $N=3$ supersymmetric charges of spinorial character, while 
for $n$: odd and $n \geq 7$ we need at least four $(N=4)$ corresponding 
quantities. In this respect, we have studied particle production in a curved 
background of five dimensions.

According to Eq. (40), the total probability to produce a particle pair over 
the entire history of the Universe is a positive, covariant quantity. In the 
spacetime under consideration, the corresponding particle production 
{\em probability density} (i.e. probability per unit of three-dimensional 
coordinate external volume) $P_2^{ext}$, may be calculated explicitly. 
Normalizing mass as 
\be
m \rarrow m_N = {m \over M_{Pl}} \leq 1 
\ee
where $M_{Pl}$ is the Planck mass, we may exclude the influence of the inner 
dimension (see also Maeda 1986). Then, the major contribution to $P_2$ comes 
out solely from the quadratic term and Eq. (40), for $j = 2$, is reduced to
\be
P_2 = {1 \over 32 \pi^2} \left [ {1 \over 90} m_N \int \: d^4 x \sqrt {-g(t)} 
\left ( {\cal R}_{\mu \nu} {\cal R}^{\mu\nu} (t) \: - \: {\cal R}_{\mu \nu 
\kp \lm} {\cal R}^{\mu \nu \kp \lm} (t) \right ) \; + \;  O (m_N^5) \right ]
\ee
which is dimensionless, as required.  The corresponding probability density 
$P_2^{ext}$, results to
\be
P_2^{ext} \simeq {1 \over 16 \pi^2} \: m_N \: \ln {t_f \over t_{in}} 
\ee
Accordingly, even at the {\em present epoch} ($t_f = t_0 = 10^{17} sec$), 
$P_2^{ext}$ represents a well-defined quantity, since
\be
P_2^{ext} \vert_{t_f = t_0} = \: 0.876 m_N \: < \: 1
\ee
Eq. (101) indicates that, the higher the masses of the created particles are, 
the greater the probability for them to be produced, will be. 
Nevertheless, as we have seen, the production of such "high-mass" particles is 
subsequently exponentially suppressed by the cosmological evolution.

From Eq. (40) [or Eq. (100)] we see that, $P_2 \neq 0$ (and therefore $ \vert 
0_{out} > \neq \vert 0_{in} >$) only as long as $m \neq 0$. For the spacetime 
(6) this is not an unexpected result, since we have considered $p_{int} = 0$. 
This condition corresponds to an adiabaticity criterion as regards the thermodynamical 
treatment of two subspaces (Kleidis and Papadopoulos 1997a) and accordingly, 
we cannot have radiation exchange (i.e. particles with $m = 0$) between them.

If we consider a quantized massive scalar field in the curved background 
under consideration, the corresponding initial and final vacuum states are 
not identical, $\vert 0_{out} > \neq \vert 0_{in} >$ and therefore an extra 
amount of particles is created in the visible space as a result of the 
semiclassical interaction between the quantum matter field and the classical 
gravitational one. Those particles correspond to field modes with 
non-vanishing momentum component in the fifth dimension $(k_5 \neq 0)$ 
(Garriga and Verdaguer 1989). The underlying mechanism of massive particle 
production in four dimensions rests in the fact that, during a period of 
high-rate cosmological contraction of the internal space, at times $t_{Pl} 
< t < t_s$, the physical frequency of the scalar modes along the fifth 
dimension is blueshifted to bring about particle creation at high energies. 
According to Eq. (66), since for $m_4 = 0$ we have $k_5 = 0$, this mechanism 
is no longer applicable to massless modes. The corresponding analysis does 
not depend on the exact value of the coupling constant involved, since, for 
the spacetime chosen, we have ${\cal R} = 0$.

One of the major problems involved in the process of particle creation in 
a curved spacetime is how to define the initial vacuum state (Hu 1974, Isham 
1981, Birrell and Davies 1982, Wald 1984). This problem is solved in the 
usual way (Hu 1974), by introducing an initial time parameter, $t_{in}$, 
prior to which the spacetime is matched to be static. This is justified by 
the fact that at the Planck epoch the semiclassical approach, which was 
taken into account, must break down. Therefore, on mathematical grounds in 
order to discuss particle creation, we need to exclude the influence 
of the period $0 < t < t_{Pl} \sim 10^{-43} sec$ and to give an unambiguous 
definition for the positive frequency modes $u_{k,k_5}^{in}$. Hence, it 
seems reasonable to assume a static {\em in-metric}. Accordingly, we assume 
that the time evolution of both subspaces starts at $t = t_{in}$. Nevertheless, 
a more natural choice of the initial vacuum is necessary. On the other hand, 
as regards the {\em out-vacuum} state (as $t \rarrow \infty$), there is no 
ambiguity, since the internal space becomes static (Fig. 1) and the external one 
exhibits adiabatic expansion, where no particles are produced (Birrell and 
Davies 1982, Garriga and Verdaguer 1989).

Once the two vacuum states ($\vert 0_{in} >$ and $\vert 0_{out} >$) are 
appropriately defined, the process of particle creation in four dimensions 
may be studied by means of the solutions of the equation governing the 
propagation of the massive quantum field in the curved spacetime (6). In this 
case, the wave equation can be solved analytically. Its exact solution has been 
obtained in terms of hypergeometric functions. Their linear transformation 
properties (Abramowitz and Stegun 1970) give the combination of positive and 
negative frequency "out-modes" which makes up a positive frequency "in-mode", 
thus indicating particle production in the out-region with respect to $\vert 
0_{in} >$. A corresponding mass-energy spectrum is attributed to the scalar 
modes as a consequence of the anisotropic evolution of the five-dimensional 
spacetime. This energy spectrum is discrete (and hence the energy is quantized) 
as a result of compactification of the extra dimension. In addition, the 
resulting energies of the field's quanta in four dimensions are strongly 
related to the inverse {\em "radius"} of the internal space, namely
$$
m_4 c^2 = \ell \times 1.977375 \times 10^{-11} {1 \over R_5} \; \; \; (MeV) 
\eqno (70)
$$
where $\ell$ is an integer and $R_5$ is measured in $cm$. In the absence of 
real experimental data regarding the extra dimensions (Barrow 1987, Casas et 
al 1987), we can only make predictions, conserning some boundary values of 
$m_4$, with respect to the corresponding values of $R_5$. 

The lower bound of $R_5$, which gives rise to the {\em upper bound} of $m_4$, 
is the Planck length (Chodos and Detweiller 1980, Kolb et al 1984, Kolb 1986, 
Accetta et al 1986, Applequist et al 1987, Barrow 1987, Garriga and Verdaguer 
1989). In this case, the energies of the particles emerging in the visible 
space are $m_4 c^2 \simeq 10^{19} GeV$, i.e. of the order of the Planck mass 
and therefore undetectable at present.

On the other hand, an upper bound for $R_5$ could be any distance which {\em 
"fits"} the radius of the internal space, provided that the extra dimension 
would remain safely {\em invisible} inside the corresponding length. In this 
case, we may consider that $R_5 \simeq 10^{-18} cm$. This distance is two 
orders of magnitude smaller than the distances at which accelerators can probe 
matter at present (Dominguez-Tenreiro and Quiros 1988, Collins et al 1989). 
Then, from Eq. (70), we obtain
\be
m_4 c^2 \simeq \ell \times 1.977 \times 10^4 \; Gev \: \simeq \: 20 \; 
TeV
\ee
thus providing a {\em lower bound} for the mass-energy of the field's 
quanta. We see that it is of the order of the estimated energy for the 
Higgs boson (Eichten et al 1984) and therefore one would expect that the 
next generation of accelerators could be able to detect such particles, 
verifying or not the matter-creation mechanism under consideration and/or 
Kaluza-Klein theories in general.  

The physical quantities corresponding to the number and the energy density 
of the created particles per unit proper volume have been explicitly 
demonstrated for the ultraviolet four-dimensional modes $(k^2 \gg 1)$, which 
correspond to particles created in the neighbouhood of the singular epoch 
$t = t_s$. It is worthnoting that, at $t = t_s$, both $\et(t_s)$ and $\rh(t_s)$ 
are well-defined and finite (see also Fig. 2)
\be
\et(t_s) = {1 \over 12 \pi^3 c^3} \; {1 \over t_s^3} \; \; \; \left ( 
{particles \over cm^3} \right )
\ee
\be 
\rh(t_s) = {3 \hbar \over 4 \pi^6 c^3} \: \zt (3) \; {1 \over t_s^4} \; \; \; 
\left ( {erg \over cm^3} \right )
\ee
From these formulas it becomes evident that the closer to the initial 
singularity the value $t = t_s$ is, the larger the amount of the produced 
particles will be.

The {\em local rate of particle production} $(LRP^2)$, per unit of proper 
three-dimensional volume and per unit time (Zel'dovich and Starobinsky 
1977, Koikawa and Yoshimura 1985) is
\be
LRP^2 = {1 \over 4 \pi^3 c^3} \: {1 \over t^4} \geq 0
\ee
Because of the spatial homogeneity of the spacetime under consideration, the 
total momentum of the quantum scalar field in the external space is conserved 
(Zel'dovich and Starobinsky 1972, Bernard and Duncan 1977, Birrell and Davies 
1892). In the case of a complex scalar field, whose quantization may be carried 
out in analogous fashion, the corresponding total charge is also conserved 
(Zel'dovich and Starobinsky 1972). Therefore, the particles are created locally, 
in pairs, with opposite charges and momenta.\\

{\bf Acknowledgements:} One of the authors (K.K.) would like to thank the 
Greek State Scholarships Foundation for the financial support during most 
of this work. The authors would like to express their gratitude to Professor 
L. Witten for his critisism and his comments on the content of this article. 
This work is partially supported by the Scientific Program PENED 1768 
(Greece).\\

\section*{References}

\begin{itemize}

\item[]Abbott R B, Barr S and Ellis S D 1984 Phys. Rev. D {\bf 30} 720
\item[]Abramowitz M and Stegun A I 1970 {\em Handbook of Mathematical 
Functions} (New York: Dover)
\item[]Accetta F, Gleiser M, Holman R and Kolb E W 1986 Nucl. Phys. B 
{\bf 276} 501
\item[]Alvarez E 1989 Rev. Mod. Phys. {\bf 61} 561
\item[]Anderson P R and Parker L 1987 Phys. Rev. D {\bf 36} 2963
\item[]Applequist T, Chodos A and Freund P 1987 {\em Modern Kaluza-Klein 
Theories} (Menlo Park: Addison - Wesley)
\item[]Bander M and Itzykson C 1966 Rev. Mod. Phys. {\bf 38} 346
\item[]Barrow J 1987 Phys. Rev. D {\bf 35} 1805
\item[]Barrow J and Ottewill A 1983 J. Phys. A {\bf 16} 2757
\item[]Bernard C and Duncan A 1977 Ann. Phys. {\bf 107} 201
\item[]Birrell N D and Davies P C W 1980 Phys. Rev. D {\bf 22} 322
\item[]Birrell N D and Davies P C W 1982 {\em Quantum Fields in Curved Space} 
(Cambridge: Cambridge University Press)
\item[]Bogolubov N N 1958 Sov. Phys. JETP {\bf 7} 51
\item[]Bunch T S, Christensen S M and Fulling S A 1978 Phys. Rev. D {\bf 18} 
4435
\item[]Bunch T S and Parker L 1979 Phys. Rev. D {\bf 20} 2499
\item[]Calan C G, Coleman S and Jackiw R 1970 Ann. Phys. {\bf 59} 42
\item[]Candelas P and Dowker J S 1979 Phys. Rev. D {\bf 19} 2902
\item[]Candelas P, Horowitz G T, Strominger A and  Witten E 1985 Nucl. Phys. 
B {\bf 258} 46
\item[]Casas J A, Martin C P and Vosmediano A H 1987 Phys Lett. B {\bf 186} 29
\item[]Chodos A and Detweiller S 1980 Phys. Rev. D {\bf 21} 2167
\item[]Christensen S M 1976 Phys. Rev. D {\bf 14} 2490
\item[]Collins P D B, Martin A D and Squires E J 1989 {\em Particle Physics 
and Cosmology} (New York: Wiley and Sons)
\item[]Davies P C W and Fulling S A 1977 Proc. R. Soc. London A {\bf 356} 237
\item[]De Witt B S 1975 Phys. Rep. {\bf 19C} 297
\item[]D' Inverno R 1993 {\em Introducing Einstein's Relativity}, pp. 284-285 
(Oxford: Oxford University Press)
\item[]Dominguez-Tenreiro R and Quiros M 1988 {\em An Introduction to 
Cosmology and Particle Physics} (Singapore: World Scientific)
\item[]Englert F 1982 Phys. Lett. B {\bf 119} 339
\item[]Eichten E, Hinchcliffe I, Lane K and Quigg C 1984 Rev. Mod. Phys. 
{\bf 56}, 579
\item[]Farina-Busto L 1988 Phys. Rev. D {\bf 38} 1741
\item[]Fischetti M V, Hartle J B and Hu B L 1979 Phys. Rev. D {\bf 20} 1757
\item[]Ford L H 1976 Phys. Rev. D {\bf 14} 3304
\item[]Forgacs P and Horvath Z 1979 Gen. Relat. Grav. {\bf 11} 205
\item[]Freund P and Rubin M 1980 Phys. Lett. B {\bf 97} 233
\item[]Friedlander F G 1975 {\em The Wave Equation on a Curved Spacetime} 
(Cambridge: Cambridge University Press)
\item[]Fulling S A 1979 Gen. Relat. Grav. {\bf 10} 807
\item[]Fulling S A, Parker L and Hu B L 1974 Phys. Rev. D {\bf 10} 3905
\item[]Garriga J and Verdaguer E 1989 Phys. Rev D {\bf 39} 1072
\item[]Gradshteyn I S and Ryzhik I M 1965 {\em Tables of Integrals Series 
and Products} (New York: Academic Press)
\item[]Green M B, Schwartz J H and Witten E 1987 {\em Superstring Theory} 
(Cambridge: Cambridge University Press)
\item[]Grib A A, Mamayev S G and Mostepanenko V M 1976 Gen. Relat. Grav. 
{\bf 7} 535
\item[]Grib A A, Mamayev S G and Mostepanenko V M 1980 J. Phys. A: Math. 
Gen. {\bf 13} 2057
\item[]Hartle J B and Hu B L 1979 Phys. Rev. D {\bf 20} 1772
\item[]Hartle J B and Hu B L 1980 Phys. Rev. D {\bf 21} 2756
\item[]Horowitz G T and Wald R M 1978 Phys. Rev. D {\bf 17} 414
\item[]Hu B L 1974 Phys. Rev. D {\bf 9} 3263
\item[]Hu B L and Parker L 1977 Phys. Lett. {\bf 63A} 217
\item[]Hu B L and Parker L 1978 Phys. Rev. D {\bf 17} 933
\item[]Isham C J 1975 {\em "An Introduction to Quantum Gravity"} in {\em 
Quantum Gravity: An Oxford Symposium} eds C J Isham, R Penrose and D W 
Sciama (Oxford: Clarendon)
\item[]Isham C J 1981 {\em "Quantum Gravity - An Overview"} in {\em Quantum 
Gravity: A Second Oxford Symposium} eds C J Isham, R Penrose and D W Sciama 
(Oxford: Clarendon)
\item[]Itzykson C and Zuber J B 1985 {\em Quantum Field Theory} (New York: 
McGraw-Hill)
\item[]Jackiw R 1974 Phys. Rev D {\bf 9} 1686
\item[]Kaluza T 1921 Sitzungsber Preuss, Akad. Wiss. Berlin, Phys. Math. 
{\bf K1} 966
\item[]Kleidis K, Varvoglis H and Papadopoulos D B 1996 J. Math. Phys. 
{\bf 37} 4025
\item[]Kleidis K and Papadopoulos D 1997{\em a} Gen. Relat. Grav. {\bf 29} 
275
\item[]Kleidis K and Papadopoulos D B 1997{\em b} J. Math. Phys. {\bf 38} 
3189
\item[]Klein O 1926 Z. Phys. {\bf 37} 895
\item[]Kobayashi S and Nomizu K 1969 {\em Foundations of Differential 
Geometry II} (New York: Wiley Interscience)
\item[]Koikawa T and Yoshimura M 1985 Phys. Lett. B {\bf 150} 107
\item[]Kolb E W, Lindley D and Seckel D 1984 Phys. Rev. D {\bf 30} 1205
\item[]Kolb E W 1986 Nucl. Phys. B {\bf 276} 501
\item[]Kolb E W and Turner M S 1990 {\em The Early Universe} (Menlo Park: 
Addison-Wesley)
\item[]Laciana C E 1993 Gen. Relat. Grav. {\bf 25} 245
\item[]Lifshitz E M and Khalatnikov I M 1963 Adv. Phys. {\bf 12} 185
\item[]Lovelock D 1971 J. Math. Phys. {\bf 12} 498
\item[]Maeda K 1984{\em a} Phys. Lett. B {\bf 138} 269
\item[]Maeda K 1984{\em b} Phys. Rev. D {\bf 30} 2482
\item[]Maeda K 1986 Phys. Lett. B {\bf 166} 56
\item[]Mashhoon B, Liu H and Wesson P S 1994 Phys. Lett. B {\bf 331} 
305
\item[]Matsuo N 1987 Z. Phys. C {\bf 36} 289
\item[]McInnes B T 1985 Phys. Lett. B {\bf 150} 113
\item[]Mijics M, Morris M S and Suen W 1986 Phys. Rev. D {\bf 34} 2934
\item[]Misner C W, Thorne K S and Wheeler J A 1973 {\em Gravitation} 
(San Francisco: Freeman)
\item[]Nesteruk A V 1991 Class. Quantum Grav. {\bf 8} L241
\item[]Nesteruk A V and Pritomanov S A 1990 Class. Quantum Grav. {\bf 7} 
L155
\item[]Parker L and Fulling S A 1974 Phys. Rev. D {\bf 9} 341
\item[]Prigogine I 1961 {\em Thermodynamics of Irreversible Processes}
(New York: Wiley)
\item[]Prigogine I, Geheniau J, Gunzig E and Nardone P 1989 Gen. Relat. 
Grav. {\bf 21} 767
\item[]Ryan M and Shepley L C 1975 {\em Homogeneous Relativistic Cosmologies} 
(New Jersey: Princeton University Press)
\item[]Ryder L H 1985 {\em Quantum Field Theory} (Cambridge: Cambridge 
University Press)
\item[]Stelle K S 1977 Phys. Rev. D {\bf 16} 953
\item[]Stelle K S 1978 Gen. Relat. Grav. {\bf 9} 353
\item[]Utiyama R and De Witt B 1962 J. Math. Phys. {\bf 3} 608
\item[]Van Nieuwenhuizen P 1977 J. Math. Phys. {\bf 18} 182
\item[]Wald R M 1984 {\em General Relativity} (Chicago: Chicago University 
Press)
\item[]Weinberg S 1979 in {\em General Relativity: An Einstein Centenary 
Survey} eds S Hawking and W Israel (Cambridge: Cambridge University Press)
\item[]Wesson P S 1992 Astrophys. J {\bf 394} 19
\item[]Zel'dovich Ya B and Starobinsky A A 1972 Sov. Phys. JETP {\bf 34} 1159
\item[]Zel'dovich Ya B and Starobinsky A A 1977 Sov. Phys. JETP Lett. {\bf 26} 
252
\item[]Zwiebach B 1985 Phys. Lett. B {\bf 156} 315

\end{itemize}

\newpage

\section*{Figure Captions}

{\bf Fig. 1:} The evolution of the internal scale factor $g_{55} (t) = 
S^2 (t)$, as a function of time. Notice that, as $t \rarrow \infty$ the 
{\em physical size} of the internal dimension, $L_5 = 2 \pi R_5 \sqrt 
{g_{55} (t)}$, is stabilized, since $S^2 (t)$ approaches to the static 
value $S^2 (t = \infty) = S_0^2$.\\

{\bf Fig. 2:} The time evolution (in dimensionless units) of the "observable" 
number (squares) and energy (asterisks) densities, for massive particles created 
in the external space. Notice the dramatic increase of both $\et_R$ and $\rh_R$ 
at the begining of the interaction procedure, to achieve their maximum values 
in the neighbourhood of the singular epoch $t_s \simeq 1.4 t_{in}$.

\end{document}